\documentclass[letter,traditabstract]{aa} 

\usepackage{longtable}
\usepackage{graphicx}
\usepackage{natbib}
\usepackage{txfonts}
\begin{document}
   \title{The long-period binary central stars of the planetary nebulae NGC~1514 and LoTr~5\thanks{Based on observations made with the Mercator Telescope, operated on the island of La Palma by the Flemish Community, at the Spanish Observatorio del Roque de los Muchachos of the Instituto de Astrof\'isica de Canarias.}}
   \titlerunning{The long-period binary central stars of the PNe NGC~1514 and LoTr~5}

   \subtitle{}

   \author{D. Jones
          \inst{1,2}
          \and
          H. Van Winckel\inst{3}
          \and
          A. Aller\inst{4}
         \and
         K. Exter\inst{3,5,6}
         \and
         O. De Marco\inst{7,8}
          }

   \institute{              Instituto de Astrof\'isica de Canarias, E-38205 La Laguna, Tenerife, Spain\\
  \email{djones@iac.es}
              \and
              Departamento de Astrof\'isica, Universidad de La Laguna, E-38206 La Laguna, Tenerife, Spain
              \and
              Instituut voor Sterrenkunde, KU Leuven, Celestijnenlaan 200D bus 2401, 3001 Leuven, Belgium\\
              \email{Hans.VanWinckel@ster.kuleuven.be}
              \and
              Instituto de F\'isica y Astronom\'ia, Facultad de Ciencias, Universidad de Valpara\'iso, Av. Gran Breta\~na 1111, 5030 Casilla, Valpara\'iso, Chile
              \and
               Herschel Science Centre, European Space Astronomy Centre, ESA, P.O.Box 78, Villanueva de la Ca\~nada, Spain
               \and
 	     ISDEFE, Beatriz de Bobadilla 3, 28040 Madrid, Spain
              \and
              Department of Physics \& Astronomy, Macquarie University, Sydney, NSW 2109, Australia
              \and
              Astronomy, Astrophysics and Astrophotonics Research Centre, Macquarie University, Sydney, NSW 2109, Australia
             }

   \date{Received February 27, 2017; accepted March 14, 2017}

 
  \abstract{The importance of long-period binaries on the formation and evolution of planetary nebulae is still rather poorly understood, in part due to the lack of central star systems known to comprise such long-period binaries.  
 Here, we report on the latest results from the on-going Mercator-HERMES survey for variability in the central stars of planetary nebulae.  
  We present a study of the central stars of \object{NGC~1514}, \object{BD+30\degr623}, the spectrum of which shows features associated with a hot nebular progenitor as well as a possible A-type companion.  Cross-correlation of high-resolution HERMES spectra against synthetic spectra shows the system to be a highly eccentric ($e\sim0.5$), double-lined binary with a period of $\sim$3300 days.   Previous studies indicated that the cool component might be a Horizontal Branch star of mass $\sim$0.55 M$_\odot$ but the observed radial velocity amplitudes rule out such a low mass. Assuming the nebular symmetry axis and binary orbital plane are perpendicular, the data are more consistent with a post-main-sequence star ascending towards the Giant Branch.  
  We also present the continued monitoring of the central star of \object{LoTr~5}, \object{HD~112313}, which has now completed one full cycle, allowing the orbital period (P$\sim$2700 days) and eccentricity ($e\sim0.3$) to be derived.  
   To date, the orbital periods of \object{BD+30\degr623} and \object{HD~112313} are the longest to have been measured spectroscopically in the central stars of planetary nebulae.  Furthermore, these systems, along with \object{BD+33\degr2642}, comprise the only spectroscopic wide-binary central stars currently known.}


   \maketitle
%

\section{Introduction}

It is now clear that a binary pathway is responsible for a significant fraction of planetary nebulae (PNe), with approximately 20\% of nebulae hosting detectable close-binary central stars \citep[CSs;][]{miszalski09a}. Space-based observations, however, have indicated that the true close-binary fraction may, in fact, be much larger \citep{demarco15}.  While the importance of such close binaries in forming axisymmetric PNe is now beyond question, the shaping influence of wider binaries \citep[which avoid the common envelope phase;][]{ivanova13,demarco09} is still uncertain.  Hydrodynamic simulations have demonstrated that the mass transfer from an Asymptotic Giant Branch (AGB) star to a companion in a wide-binary can be, under certain circumstances, greatly enhanced with respect to the simple Bondi-Hoyle-Lyttleton accretion rate with significant material also deposited into the orbital plane of the binary \citep{theuns96}.  Even if the right conditions do not arise and the accretion rate onto the companion is modest, simulations show that the focussing of the AGB wind by the companion should still cause a departure from spherical symmetry \citep[e.g.][]{edgar08}.  The results of these simulations would seem to indicate that wide-binaries can also have a significant influence on the morphology of any resulting PN, however there are so few PNe known to harbour wide binaries that it is particularly difficult to characterise their influence observationally.

Several extremely wide, visual binaries have been discovered using the Hubble Space Telescope \citep{ciardullo99}, however these systems are so wide that it is highly unlikely that there is significant interaction between the components of the central binary.  A few wide systems have been discovered through the composite nature of the spectra of their CSs \citep[e.g.\ 
\object{Hen~2-172};][]{
pereira10}, many of which show companion stars which have been contaminated by material transferred from the nebular progenitor while it was on the AGB \citep[known as Barium stars due to the prominent presence of Barium lines in the spectrum of the contaminated companion; ][]{bond03,miszalski12a,miszalski13a}.  The morphologies of these systems do, indeed, seem to indicate that such wide binaries may have an important shaping influence on their host PNe, with a prevalence for dense equatorial rings and bipolar structures \citep{tyndall13} similar to those seen in the close binary population \citep{miszalski09b}.  Based on studies of naked Barium stars (i.e.\ those which do not have PNe around them), these systems are expected to present orbital periods roughly in the range of 100-10\,000 days \citep[e.g.][and references therein]{vanderswaelmen17}.  However, to date, only one wide-binary CS, \object{BD+33\degr2642} (\object{PN~G052.7+50.7}), has a confirmed orbital period (at 1105 $\pm$ 24 days) and, intriguingly, is not a Barium star \citep{vanwinckel14}.  \cite{demarco04} found that 10 out of a sample of 11 CSs (including \object{BD+33\degr{}2642}) monitored spectroscopically showed variable radial velocities, suggesting that there may be a significant population of, as yet undiscovered, long-period binary CSs.

Here, we present the results of the long-term radial velocity monitoring programme of PN CSs responsible for determining the orbital period of \object{BD+33\degr2642}.  The survey employs the HERMES spectrograph \citep{raskin11} mounted on the 1.2 m Mercator Telescope at the Observatorio del Roque de Los Muchachos on the Spanish island of La Palma.  The programme has been running since the installation of HERMES in 2009 and takes advantage of the instrument's high throughput, stability and spectral resolution to search for small amplitude radial velocity variations associated with long-period variables \citep{vanwinckel15}.  Further details of the instrument operation and standardised data reduction for the programme can be found in \cite{raskin11} and \cite{vanwinckel14}.

This letter is organised as follows: Section \ref{sec:ngc1514} introduces \object{NGC~1514} and the results related to its CS are found in Section \ref{sec:results}.  In Section \ref{sec:update}, we present an update of the results presented in \cite{vanwinckel14} on \object{LoTr~5}, before concluding in Section \ref{sec:conclusions}.

\section{NGC~1514}
\label{sec:ngc1514}

\begin{figure}[]
\centering
\includegraphics[width=0.8\columnwidth,angle=270]{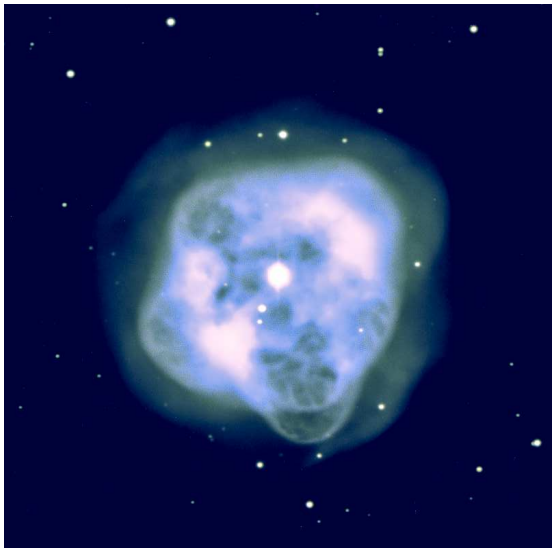}
\caption[]{An image of \object{NGC1514} produced using archival O[\textsc{iii}] and H$\alpha$+N[\textsc{ii}] images taken with the Isaac Newton Telescope's Wide Field Camera.  The image measures 4.5\arcmin$\times$4.5\arcmin{}, North is up and East is left.}
\label{fig:image}
\end{figure}

 \begin{figure}[]
\centering
\includegraphics[width=1.0\columnwidth,angle=0]{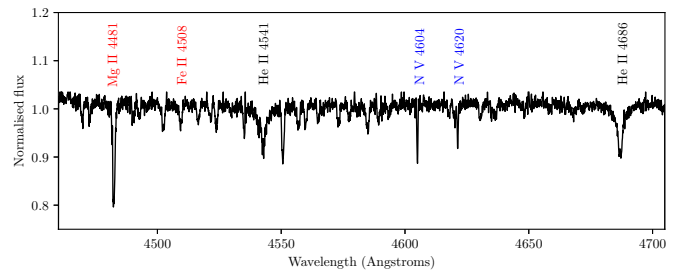}
\caption[]{An example subsection of a HERMES spectrum of \object{BD+30\degr623} with selected features from both hot (blue) and cool (red) components marked.  Marked in black are He~\textsc{ii} absorption features that may contain contributions from both components and were therefore masked off in the analysis.}
\label{fig:1514spec}
\end{figure}

The nebula \object{NGC~1514} itself presents a highly complex structure with multiple shells \citep[see figure \ref{fig:image};][]{chu87}.  The outer shell appears more or less elliptical, while the inner region has a more complicated structure that has been interpreted as being ``composed of numerous small bubbles'' \citep{hajian97}.  WISE imagery of the nebula revealed the presence of a pair of axisymmetric rings dominated by dust emission \citep{ressler10}, bearing striking resemblance to the symbiotic nebula \object{Hen~2-104} \citep[the Southern Crab;][]{santander08}.  Such structures are considered to be a natural product of binary interactions \citep{sahai99,soker00}, leading \cite{ressler10} to conclude that \object{NGC~1514} must have been formed by an interacting binary system.

The CS of \object{NGC~1514}, \object{BD+30\degr623}, was first identified as displaying a composite spectrum by \cite{kohoutek67}, with the analysis of \cite{kohoutek67b} finding the two components to be an A-type main sequence star and a blue subdwarf (sdO, assumed to be the nebular progenitor). The more detailed analysis of \cite{aller15} found the spectrum of the hot component to be consistent with an sdO classification and the cool component to be an A0-type star likely in a Horizontal Branch (HB) evolutionary phase, although the possibility that the star is a more massive post-main-sequence (post-MS) star could not be ruled out \citep[see also][]{mendez16}.  \cite{montez15} showed the CS to be a compact source of X-rays, the potential origin of which they were unable to constrain.  However, the emission could arise from coronal activity of the secondary, more consistent with a post-MS evolutionary status.
 
Repeated efforts to monitor the CS system, both photometrically and spectroscopically, for signs of variability associated with binarity have not borne fruit - most conclusively \cite{mendez16} obtained high-resolution spectra over a time-span of 500 days showing no appreciable variation in radial velocity for either component.  Furthermore, in addition to being stable over the course of the observations, the two components presented radial velocities which differ from one another by approximately 13 km~s$^{-1}$, where the radial velocity of the hot component is consistent with the weighted mean of literature measurements for the radial velocity of the nebula \citep[there is significant variation in values in the literature, possibly indicating that this may not be completely reliable;][]{schneider83}. As a result, \cite{mendez16} concluded that the two components are, in fact, not related and merely a near-perfect chance alignment.

\section{Analysis of NGC1514}
\label{sec:results}

\begin{figure}[]
\centering
\includegraphics[width=\columnwidth,angle=0]{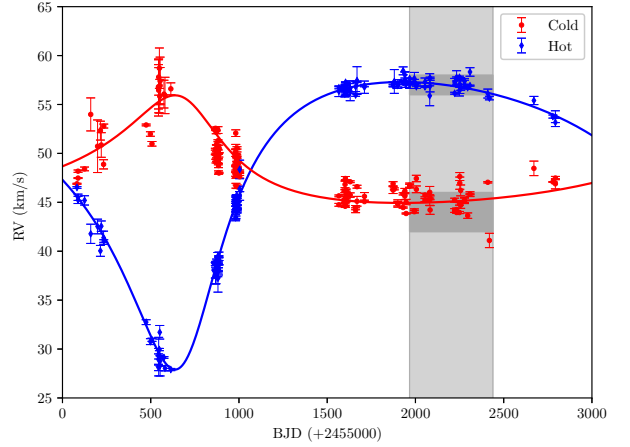}
\caption[]{Radial velocity curves of the hot and cool components of \object{BD+30\degr623}.  The light grey shaded region marks the time coverage of the study of \citet{mendez16}, while the dark grey regions enclose their confidence limits for the radial velocities of the two components.}
\label{fig:1514rv}
\end{figure}

Radial velocities of the two components of the CS of \object{NGC~1514}, \object{BD+30\degr623}, were derived via cross-correlation of 143 spectra (an example of which is shown in Figure \ref{fig:1514spec}), obtained over a time-span of nearly 2711 days, with synthetic spectra produced individually for each component.  For the cool component, the synthetic spectrum fit from \cite{aller15} was used, while for the hot component a synthetic spectrum with identical parameters to those of \cite{aller15}, but with solar abundances (their spectrum includes only H and He), was produced using the TMAP package \citep{werner03}.  Regions containing lines with strong nebular contamination (i.e.\ Balmer lines, O[\textsc{iii}] at 4959 and 5007\AA{}) or which could present contributions from either component (i.e.\ He~\textsc{ii} 4541\AA{}) were masked off in order to avoid contamination in the resulting cross-correlation functions.  The resulting radial velocity variations are plotted in Figure \ref{fig:1514rv}, clearly demonstrating a long-term variability due to a highly eccentric binary.  From the figure, it is easy to see why \cite{mendez16} found no evidence of orbital motion in their data given the change in radial velocity for both components is of order 1~km~s$^{-1}$ during time-span of their observations (grey shaded area in Figure \ref{fig:1514rv}).  Indeed, over those dates, our observations agree well with the confidence ranges (the dark grey shaded areas in Figure \ref{fig:1514rv}) for the the radial velocities of each component given by \cite{mendez16}.

An orbital solution was derived through simultaneous fitting of both radial velocity curves, with the resulting parameters listed in Table \ref{tab:ngc1514}.  The fits are also shown overlaid on the radial velocity data in Figure \ref{fig:1514rv}, from which one can appreciate the quality of the fit to the hot component.  The cool component, however, is not so well fit.  Fitting to the measured radial velocities of the cool component alone results in a greater eccentricity value ($\sim$0.8), lower systemic velocity ($\sim$48 km~s$^{-1}$) and larger amplitude ($\sim$7 km~s$^{-1}$) compared to those values determined by simultaneous fitting of both curves.  However, the overall fit to the cool component's radial velocity curve does not improve greatly and the values listed in Table \ref{tab:ngc1514} lie comfortably within 1$\sigma$ of the values derived from fitting the cool component independently.  The poor fit and general scatter 
of the points is almost certainly attributable to the rotational velocity of the star which, while not extreme \cite[$\sim$40 km~s$^{-1}$; ][]{greenstein72}, is much larger than the amplitude of the orbital motion.

\cite{aller15} found that the cool A-type component could be either on the HB with mass $\sim$0.55 M$_\odot$ or, alternatively, a post-MS star with mass $\sim$3 M$_\odot$.  While the observed radial velocity amplitude ratio (and therefore mass ratio) of the SB2 binary has a large uncertainty, mainly due to the large scatter in the velocities of the cool component, we can safely state that the mass ratio, q, is $\gtrsim$2.  Such a mass ratio effectively rules out the HB scenario irrespective of the possible mass of the hot component.  However, the masses of both components can be estimated by assuming that the orbital plane lies perpendicular to the nebular symmetry axis, as predicted by PN shaping models and observed in close binary CSs \cite{hillwig16}.  \cite{ressler10} determine a nebular inclination of 59\degr{} (i.e.\ a 31\degr{} inclination for the binary) which, when combined with the radial velocity amplitudes and measured eccentricity, implies a mass of 2.3$\pm$0.8 M$_\odot$ for the cool secondary (not consistent with the HB scenario).  The mass for the hot primary is found to be 0.9$\pm$0.7 M$_\odot$, greater than a typical sdO star mass but still well within 1$\sigma$.

\begin{table}
\caption{Table of fitted parameters and uncertainties (derived using a Markov Chain Monte Carlo process) for \object{BD+30\degr623} (\object{NGC~1514})}             
\label{tab:ngc1514}      
\centering                         
\begin{tabular}{c c c}        
\hline\hline                
Parameter & Hot & Cool \\    
\hline                        
K (km~s$^{-1}$) & 14.7$\pm$0.1 & 5.5$\pm$2.0\\
Period (days) & \multicolumn{2}{c}{3306$\pm$60}\\
Eccentricity & \multicolumn{2}{c}{0.46$\pm$0.11} \\
Systemic velocity, $\gamma$ (km~s$^{-1}$) & \multicolumn{2}{c}{48.7$\pm$0.5}\\
q$\equiv$$\frac{\mathrm{M}_\mathrm{Cool}}{\mathrm{M}_\mathrm{Hot}}$ & \multicolumn{2}{c}{2.7$^{+1.6}_{-0.7}$}\\
\hline                                   
\end{tabular}
\end{table}

\section{Update on LoTr~5}
\label{sec:update}
\begin{figure}[]
\centering
\includegraphics[width=\columnwidth,angle=0]{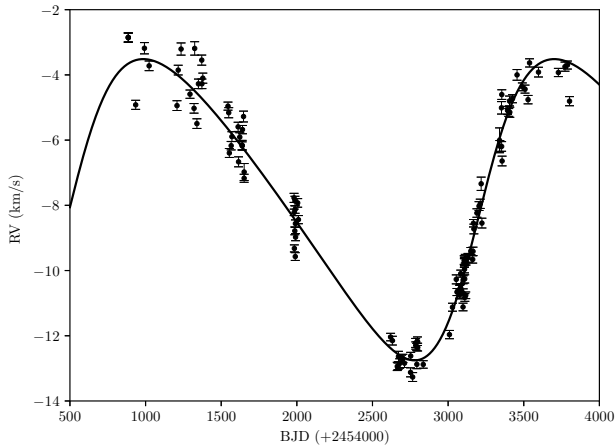}
\caption[]{Radial velocity curve of \object{HD~112313} (PN LoTr~5).}
\label{fig:lotr5rv}
\end{figure}

\cite{vanwinckel14} presented the first detections of orbital motion due to a long-period binary CS, with the detections of variability in \object{BD+33\degr2642} and \object{LoTr~5}. At that time the data for \object{LoTr~5}
 showed only a slow decline of $\sim$10 km~s$^{-1}$ over a time-span of 1807 days.  Monitoring of the CS of \object{LoTr~5} (\object{HD~112313}) has continued as part of the Mercator-HERMES programme and has now covered an entire orbital period.  The now-complete radial velocity curve, obtained using the same methodology as that of \cite{vanwinckel14}, is shown in Figure \ref{fig:lotr5rv} along with a fit to the orbital solution (the parameters of which are listed in Table \ref{tab:lotr5}).  
 We do not find any evidence of the second component reported by \citet{jasniewicz87}, neither directly  in our spectra or at the reported velocity amplitudes (K=40 km~s$^{-1}$) in our cross-correlation functions.

The primary mass function, $f(m_1)$, of HD~112313 is found to be 0.025$\pm$0.003 M$_\odot$,
 lower than that estimated by \citet[][as their estimation assumed a circular orbit]{vanwinckel14}.  However, the apparent inconsistency with the nebular inclination as derived by \cite{graham04} still persists given that, at that inclination ($i=17\degr$) and for any reasonable value of the mass of the G-type secondary component (i.e. 1 M$_\odot$), the mass of the primary component would still be in excess of 2 M$_\odot$.  As the central white dwarf cannot have such a high mass (greater than the Chandrasekhar mass), only a few conclusions are possible:
 \begin{itemize}
 \item The inclination derived by \cite{graham04} is incorrect.  As the determination of primary mass is highly dependent on the inclination (to the third power), a small error in the inclination could resolve the issue (for example, at inclinations $>$23\degr{} the mass of the primary would be $<$1.4 M$_\odot$).
 \item The nebular waist is not coplanar with the binary plane, at odds with the results from PNe with close-binary CSs which are found to always be coplanar \citep{hillwig16}.
 \item The system is a hierarchical triple where the central white dwarf actually forms part of a close binary, the total mass of which is $\gtrsim$2 M$_\odot$.  The spectral non-detection of a solar mass close-binary companion as well as stability considerations throughout the system's evolution question the likelihood of this conclusion. \end{itemize}

\begin{table}
\caption{Table of fitted parameters and uncertainties (derived using a Markov Chain Monte Carlo process) for \object{HD~112313} (\object{LoTr~5})}             
\label{tab:lotr5}      
\centering                         
\begin{tabular}{c c}        
\hline\hline                
Parameter & Value \\    
\hline                        
K (km~s$^{-1}$) & 4.6$\pm$0.1\\
Period (days) & 2717$\pm$63\\
Eccentricity & 0.26$\pm$0.02 \\
Systemic velocity, $\gamma$ (km~s$^{-1}$) & $-$7.9$\pm$0.1\\
\hline                                   
\end{tabular}
\end{table}

\section{Conclusions}
\label{sec:conclusions}

We present derivations of the orbital periods of the CSs of \object{NGC~1514} and \object{LoTr~5} \citep[the latter of which was shown to be variable in a previous paper;][]{vanwinckel14} which are now the longest known orbital period CSs, to date.

The CS of \object{NGC~1514}, \object{BD+30\degr623}, has long been known to present a composite spectrum with features attributable to a hot sdO star (assumed to be the nebular progenitor) and a cooler A0-type star.  Previously radial velocity studies, including the long-term ($\sim$ 1 year) monitoring of \cite{mendez16}, failed to find clear signs of variability, with some authors even going so far as to claim that the two stars are unrelated \citep{mendez16}.  Given the $\sim$9 year orbital period and high ($e\sim0.5$) eccentricity, it is no surprise that previous studies have failed to see signs of orbital motion, let alone derive the orbital period - highlighting the importance of extremely long-term ($\sim$years) monitoring programmes \citep{vanwinckel15}.  Previous studies of the system presented two different scenarios for the A0 component of the composite spectrum, a low mass HB star or a more massive post-MS star \citep{aller15}.  The amplitudes of the radial velocity curves rule out the HB scenario, while adopting the nebular inclination provides rough mass estimates for both components of M$_\mathrm{Hot}\sim$0.9 M$_\odot$ and M$_\mathrm{Cool}\sim$2.3 M$_\odot$.

The CS of \object{LoTr~5}, \object{HD~112313}, was demonstrated in a previous paper to show radial velocity variability associated with binarity \citep{vanwinckel14}, but a full orbit had not yet been covered.  With the addition of new data, the full orbit has now been sampled allowing the derivation of the orbital parameters, showing HD~112313 to have an eccentricity of $\sim$0.3 and an orbital period of $\sim$8 years.  Providing the inclination is indeed 17\degr{} \citep[i.e.\ perpendicular to the nebular symmetry axis;][]{graham04}, the mass function of the primary implies an impossibly high white dwarf mass ($>$1.4 M$_\odot$).  This may be an indication that the binary system is not coplanar with the waist of the nebula or, perhaps more likely, the previous determination of the nebular inclination is incorrect.  Alternatively, the system may be a hierarchical triple where the observed G-type star is in wide orbit around a binary system comprised of the nebular progenitor and a close companion of mass $\sim$1.5 M$_\odot$.  We encourage further study of both the nebula and its CS to evaluate these scenarios.

\begin{acknowledgements}
Based on observations obtained with the HERMES spectrograph, which is supported by the Fund for Scientific Research of Flanders (FWO), Belgium, the Research Council of KU Leuven, Belgium, the Fonds National de la Recherche Scientifique (FNRS), Belgium, the Royal Observatory of Belgium, the Observatoire de Gen\`eve, Switzerland and the Th\"uringer Landessternwarte Tautenburg, Germany. The Mercator telescope is operated thanks to grant number G.0C31.13 of the FWO under the ``Big Science" initiative of the Flemish government. H.V.W. acknowledges support from The Research Council of the KU Leuven under grant number GOA/2013/012.  A.A. acknowledges support from FONDECYT through postdoctoral grant 3160364. The authors want to thank all observers of the HERMES consortium institutes (KU Leuven, ULB, Royal Observatory, Belgium, and Sternwarte Tautenburg, Germany), who contributed to this monitoring programme.  This paper makes use of data obtained from the Isaac Newton Group Archive (maintained as part of the CASU Astronomical Data Centre at the Institute of Astronomy, Cambridge).  The TMAW tool (http://astro.uni-tuebingen.de/~TMAW) was constructed as part of the activities of the German Astrophysical Virtual Observatory.
\end{acknowledgements}

\bibliographystyle{aa}
\bibliography{literature}

\end{document}